# Kane-Pumplin-Repko Factorization: its application to precision measurements of transverse spin asymmetries and to the study of TMD evolution.

## Dennis Sivers

Portland Physics Institute
4730 SW Macadam #101
Portland, OR 97239

University of Michigan
Spin Physics Center
2477 Randall Lab
Ann Arbor, MI 48109

## Abstract

This article presents a summary of overlapping presentations by the author to the *QCD Evolution 2013 Workshop* (Jefferson Lab, May 6-10, 2013) and to the *Opportunities for Polarized Physics at Fermilab* workshop (Fermilab, May 20-22, 2013). It contains an introduction to the concept of Kane-Pumplin-Repko (KPR) factorization and describes how this concept can be used in the analysis of high precision measurements of parity-conserving transverse single-spin asymmetries. The discussion demonstrates that such measurements can not only probe directly for specific mechanisms that enhance our fundamental understanding of nonperturbative QCD dynamics but, because transverse spin asymmetries are unambiguously parameterized by a spin-directed momentum shift, $\langle \delta k_{TN}(x,\mu^2) \rangle$, such measurements can also be used to calibrate other phenomenological applications of transverse momentum dependent distributions (TMD's) and of TMD evolution. The calibration supplied by these measurements can thus enable the use of TMD factorization for the exploration of a broad range of other aspects of hadronic structure. KPR factorization ensures that $\langle \delta k_{TN}(x,\mu^2) \rangle$ remains invariant under TMD evolution and this invariance can be used in the precision comparison of transverse single-spin asymmetries in the Drell-Yan process with those in Semi-inclusive deep inelastic scattering.

PACS  11.15.-q-   gauge field theories
PACS  11.30.Rd   chiral symmetries
PACS  12.38.Aw   general properties of QCD(dynamics of confinement, etc.)
PACS  13.88.+e   polarization in interactions and scattering

July 20, 2013

# 1. Introduction to Kane-Pumplin-Repko Factorization—symmetries, idempotent projection operators and superselection in quantum field theory.

In "The Adventure and the Prize" [1] the author discussed the close connection found in non-perturbative QCD between transverse single-spin observables and the complex dynamics of color confinement combined with the dynamical breaking of chiral symmetry. Since the publication of that paper, there have been significant developments that suggest it is appropriate to revisit in more detail the implications of KPR factorization for the high-precision measurements presented there. One notable development consists of the stage-1 approval for a measurement of $A_N d\sigma(p\uparrow p \Rightarrow \mu^+\mu^-(M^2)X)$ in the SeaQuest detector at Fermilab (E-1027) [2]. This approval combined with the extensive program for the measurement of transverse spin asymmetries in semi-inclusive deep inelastic scattering (SIDIS) processes that has been approved for early measurements with the JLAB 12-GeV upgrade [3] now allows for the timely execution of precision tests for "Collins Conjugation" [4]. This term describes the prediction that relates orbital distributions leading to transverse spin asymmetries measured in the Drell-Yan process to those measured in SIDIS, by

$$\Delta^N G^{eff}_{q/p\uparrow}(x, k_{TN}, \mu^2)_{DY} = -\Delta^N G^{eff}_{q/p\uparrow}(x, k_{TN}, \mu^2)_{SIDIS} \tag{1.1}$$

Direct experimental comparisons testing this prediction have been identified as a high priority goal in hadron physics [5]. Precision tests of Collins Conjugation were also a major theme in both the *QCD Evolution 2013 Workshop* and the *Opportunities for Polarized Physics at Fermilab* workshop. Definitive experimental tests of (1.1) involve applications of the theoretical and phenomenological advances in the understanding of TMD evolution. Both the CSS [6] and EIS [7] formalisms are now up to the challenge of confirming or refuting this prediction. The technical aspects of TMD evolution, however, will not be discussed here. Instead, we will concentrate on the tools, including projection operators and superselection, brought into calculations and predictions of transverse single-spin asymmetries by Kane-Pumplin-Repko factorization.

The precise formulation of KPR factorization is based on the result of a calculation that can be found in Ref. [8],

$$A_N[q_1(p_1)q_2\uparrow(p_2) \Rightarrow q_f(k_f)\mu^2(s)] = \frac{d\sigma(qq\uparrow \Rightarrow q) - d\sigma(qq\downarrow \Rightarrow q)}{d\sigma(qq\uparrow \Rightarrow q) + d\sigma(qq\downarrow \Rightarrow q)} \tag{1.2}$$

$$= \alpha_s(s)\frac{m_q}{\sqrt{s}} f_{TN}(\Theta_{CM})(1+O(\alpha_s))$$

In this calculation $s = (p_1 + p_2)^2 = 4[\vec{k}_f^{\,2} + \frac{1}{2}(\mu^2(s) + m_q^2)]$ is the s-channel Mandelstam invariant of the hard collision, and,

$$f_{TN}(\Theta_{CM}) = \frac{k_{TN}}{(\mu^2(s) + k_{TN}^2 + k_{TS}^2)^{\frac{1}{2}}} \tilde{f}(\Theta_{CM}) \qquad (1.3)$$

is an angular function that vanishes at $\Theta_{CM} = 0$. Equation (1.3) introduces the definition of a spin-directed transverse momentum

$$k_{TN} = \vec{k}_f \cdot (\hat{s} \times \hat{p}_1), \qquad (1.4)$$

that is characteristic of all parity-even single-spin asymmetries. Spin-directed transverse-momentum asymmetries of this type will play an important role in this discussion. Asymmetries in the remaining orthogonal component of transverse momentum parallel to the spin direction, here designated $k_{TS}$, are odd under both parity and time-reversal and are highly suppressed in all scattering processes. (The Trento conventions [9] used to determine the experimental definitions applicable to such measurements contain a more complete set of definitions)

The authors of ref. [8] interpreted their calculation (1.2) within the framework of collinear factorization for hard processes and suggested in the conclusion of the paper that transverse single-spin asymmetries would vanish at large values of $\sqrt{s}$. For several years, the suggestion was quoted [10] as a prediction of QCD. However, it is now apparent that this original interpretation was flawed. While subsequent studies confirmed that the KPR result (1.2) was not enhanced in higher orders of perturbation theory, Efremov and Teryaev [11] correctly argued that the KPR calculation was just one example of a twist-3 observable, one suppressed by a particularly unfavorable factor of $m_q/\sqrt{s}$, and proposed that other twist-3 operators with less suppression should instead be considered in the analysis of transverse spin asymmetries. In addition, a different interpretation of the KPR calculation was presented that leads to the following definition:

**Kane-Pumplin-Repko factorization in QCD: In the limit $m_q \to 0$ there exists within QCD perturbation theory a symmetry that is strongly broken in the nonperturbative sector of the theory. The existence of this symmetry ensures that <u>all</u> parity-even single spin asymmetries in hard-scattering processes can be absorbed into the transverse-momentum dependence of hadronic distribution functions or fragmentation functions. [12,13]**

Although the definition refers to transverse-momentum dependence of hadronic distribution functions or fragmentation functions, KPR factorization is quite distinct from TMD factorization.[6,7] In fact, KPR factorization applies in processes in which there may not be TMD factorization but in which factorization still exists at the level of $k_T$-integrated, collinear, distributions or fragmentations functions. This is because KPR factorization encompasses the full range of single-spin asymmetries that can be generated by higher-twist dynamical mechanisms, including those mechanisms discussed by Efremov and Teryaev [11], Qiu and Sterman [14], as well as those involving fragmentation processes presented by Metz and Pitonyak[15]. Indeed, KPR factorization allows for the normalization of the specific operators included in these higher-twist calculations within the framework of nonperturbative spin-orbit dynamics and the isolation of spin effects implied by KPR factorization means that the process dependence of orbital distributions and Boer-Mulders functions can eventually be understood within the extended universality of higher-twist effects found in collinear factorization. Theoretical challenges remain in the interpretation of these functions and their applications to processes without TMD factorization but it is certain that KPR factorization can help define the boundaries of the relevant phenomenology. In processes, such as DY and SIDIS, where TMD factorization is known to be valid, KPR factorization organizes and isolates the spin dependence of the calculations. As can be seen from the definition, strict application of KPR factorization requires that perturbative QCD calculations be executed with $m_q = 0$. This application can thus be justified for spin asymmetries in hard-scattering processes involving u or d quarks and, with some attention to possible corrections [16] to processes involving s quarks. In addition, the quark-mass dependent terms of single-spin asymmetries involving heavy (c,b,t) quarks can be calculated in QCD perturbation theory. However, it may be that spin-orbit dynamics introduces nonperturbative corrections even in these cases. In what follows, we will sometimes use the term "KPR isolation" instead of "KPR factorization" to emphasize both its distinction from and its compatibility with TMD factorization.

The arguments in Ref. [1] identify the underlying chiral invariance of perturbative QCD with $m_q = 0$ as the origin of KPR factorization. However, in order to effectively use the concept in quantum field-theoretical calculations, it is necessary to more accurately describe the specific symmetry involved. To this end, it is helpful to observe that all single-spin asymmetries can be described as observables of the form,

$$A_\sigma(\{\vec{k}_i\};\vec{\sigma}) = \frac{N(\{\vec{k}_i\};\vec{\sigma}) - N(\{\vec{k}_i\};-\vec{\sigma})}{N(\{\vec{k}_i\};\vec{\sigma}) + N(\{\vec{k}_i\};-\vec{\sigma})} \qquad (1.5)$$

where the measured spins, $\pm\vec{\sigma}$, are axial 3-vectors and the kinematics are specified by a set, $\{\vec{k}_i\}$, of momentum 3-vectors. All such observables are therefore <u>odd</u> under an operator, O, that acts on a set,

$$\{\vec{k}_i;\vec{\sigma}_j\} = \{\vec{k}_i\} \cup \{\vec{\sigma}_j\} \qquad (1.6)$$

of 3-vectors $\{k_i\}$ and axial 3-vectors $\{\vec{\sigma}_j\}$ such that:

$$O\{\vec{k}_i;\vec{\sigma}_j\}O^{-1} = \{\vec{k}_i;-\vec{\sigma}_j\}. \qquad (1.7)$$

The operator, O, has been designated [17] the "snake operator" in recognition of the Siberian snake of accelerator physics invented by Derbenev and Kondratenko [18].

The snake operator serves as a 3-dimensional Hodge dual form of the familiar parity operator, P,

$$\mathrm{P}\{\vec{k}_i; \vec{\sigma}_j\} P^{-1} = \{-\vec{k}_i; \vec{\sigma}_j\} \tag{1.8}$$

The operator that defines the symmetry involved in the definition of KPR factorization is then the product of these two operators, $\mathrm{A}_\tau = \mathrm{OP}$. The operator $\mathrm{A}_\tau$ is sometimes designated "naïve time reversal" [19] or "artificial time reversal" [20] and can be seen to have the action;

$$\mathrm{A}_\tau \{\vec{k}_i; \vec{\sigma}_j\} \mathrm{A}_\tau^{-1} = \{-\vec{k}_i; -\vec{\sigma}_j\} \tag{1.9}$$

The scalar observable, $k_{TN}$, defined in Eqs. (1.3) – (1.4) can be seen to be an $\mathrm{A}_\tau$-odd quantity. As we shall describe later, the operator $\mathrm{A}_\tau$ defines the specific symmetry in KPR factorization that is preserved in QCD perturbation theory with $m_q = 0$ but is broken by the non-local spin-orbit dynamics found in confined or strongly-interacting systems.

The operators $\mathrm{P}, \mathrm{O}, \mathrm{A}_\tau$ have a group structure defined by $\mathrm{PO} = \mathrm{A}_\tau, \mathrm{OA}_\tau = \mathrm{P}, \mathrm{A}_\tau \mathrm{P} = \mathrm{O}$ with $\mathrm{P}^2 = \mathrm{O}^2 = \mathrm{A}_\tau^2 = \mathrm{POA}_\tau = 1$ and each of these operators can be used to define quantum-mechanical idempotent projection operators. The group structure leads to the classification of all single-spin observables of the form (1.5) into two categories:
      1. $\mathrm{P}$-odd and $\mathrm{A}_\tau$-even
      2. $\mathrm{A}_\tau$-odd and $\mathrm{P}$-even.

In the standard model of particle physics, $\mathrm{P}$-odd, longitudinal single-spin asymmetries are associated with the $W^\pm, Z_0$ interactions of electroweak theory while the $\mathrm{A}_\tau$-odd, transverse asymmetries involve either quark masses or nonperturbative dynamical mechanisms that break the chiral invariance of the quark sector Lagrangian of QCD. Because $\mathrm{A}_\tau$-odd dynamics can involve non-local spin-orbit correlations there exist process dependence in $\mathrm{A}_\tau$-odd observables. One consequence of the group structure is that parity-odd single-spin asymmetries involving hadrons inherit a process dependence engendered by the requirement that such asymmetries must also be $\mathrm{A}_\tau$-even. This process dependence must be considered when using spin-observables to probe for dynamics beyond the standard model with hadronic systems.

The operator $\mathrm{A}_\tau$ has occasionally been confused with the actual time-reversal operator, T, of quantum field theory. In order to appreciate the application of KPR isolation or KPR factorization, it is convenient to keep in mind the following distinctions:

1. $A_\tau$ is unitary while T is antiunitary and, in scattering processes, involves the interchange of initial and final states and re-ordering of operators.
2. $A_\tau^2 = 1$ while $T^2 = (-1)^F$ where F is the fermion number operator. This distinction is important in understanding the connection between spin and statistics in quantum field theory.
3. $A_\tau$-effects in QCD are very large compared to T-odd effects.
4. Spin-orbit effects are T-even and the transverse momentum observable $k_{TN}$ defined in (1.3)-(1.4) is $A_\tau$-odd and T-even.

The fourth point is somewhat redundant but is included here because it is still sometimes misstated in arguments and calculations. In the next section, we will use the power of the projection operators,

$$\Pi_A^\pm = \frac{(1 \pm A_\tau)}{2} \qquad (1.10)$$

to isolate the $A_\tau$-odd dynamics of spin-orbit correlations from the $A_\tau$-even dynamics involved in the hard scattering of a massless quark in order to define an orbital distribution function. Using (1.9), the operator $A_\tau$ can be represented by $\hat{\Sigma}_y$ which projects angular momentum normal to the x-z plane. This representation is, of course, basis dependent. In the transversity basis for a spin-$\frac{1}{2}$ particle:

$$A_\tau^{transv} = \begin{pmatrix} 1 & 0 \\ 0 & -1 \end{pmatrix} \qquad (1.11)$$

While in the helicity basis:

$$A_\tau^{hel} = \begin{pmatrix} 0 & i \\ -i & 0 \end{pmatrix} \qquad (1.12)$$

The basis dependence leads to two equivalent projections. In the transversity basis we have:

$$|M|^2 \uparrow = |M^+|^2 + |M^-|^2$$
$$|M|^2 \downarrow = |M^+|^2 - |M^-|^2 \qquad (1.13)$$
$$A_N = |M^-|^2 / |M^+|^2$$

while in the helicity basis we write:

$$|M|^2 \uparrow = |M_0|^2 + |F|^2 + \text{Im}(FM_0^*)$$
$$|M|^2 \downarrow = |M_0|^2 + |F|^2 - \text{Im}(FM_0^*) \qquad (1.14)$$
$$A_N = \text{Im}(FM_0^*) / (|M_0|^2 + |F|^2)$$

Feynman diagram calculations require the rotational invariance found in helicity amplitudes and Feynman diagram calculations of $A_\tau$-odd effects often use the imaginary part of an individual diagram without explicitly verifying that it represents a spin-flip projection in the x-z plane. In addition, Feynman diagram calculations of $A_\tau$-odd dynamics often calculate $M_0$ and $F$ in different orders of perturbation theory and thus do not include both terms in the denominator of (1.14). The recent paper by Broksky, Hwang, Kovchegov, Schmidt and Sievert [21] discusses the analysis of such Feynman diagram calculations in more detail.

## 2. The $A_\tau$-odd dynamics of the pion tornado in the Georgi- Manohar Chiral Quark Model.

In order consider the implications of KPR factorization it is convenient to consider the application of the projection operators in the context of an explicit model. Therefore, we are going to illustrate the isolation of spin-orbit dynamics within a simplified version of the Georgi-Manohar [22] chiral quark model. We will start by considering a proton with spin oriented in the $+\hat{y}$ direction and restrict attention to a configuration consisting of an $I=0, J^P = 0^+$ constituent diquark $[U,D]$ bound to a constituent quark $U\uparrow$ with spin also oriented in the $+\hat{y}$ direction. For convenience, at this initial stage we do not consider any orbital angular momentum in the diquark-quark system. We can designate the configuration as

$$\left|p\uparrow; P_p^\nu\right\rangle \cong [U,D](P_0^\nu) \otimes \left|U\uparrow; P_U^\nu\right\rangle \qquad (2.1)$$

For convenience, we express the 4-momenta in light-cone coordinates with

$$P_p^\nu = P_0^\nu + P_U^\nu$$
$$P_p^\nu = (P^+, \frac{m_p^2}{P^+}, 0, 0)$$
$$P_0^\nu = ((1-x)P^+, \frac{m_0^2(eff) + p_x^2 + p_y^2}{(1-x)P^+}, -p_x, -p_y) \qquad (2.2)$$
$$P_U^\nu = (xP^+, \frac{m_U^2(eff) + p_x^2 + p_y^2}{xP^+}, p_x, p_y)$$

In these expressions, the longitudinal, $p_z$, motion of the $[U,D]$ diquark and constituent U quark are absorbed into $m_0^2(eff)$ and $m_U^2(eff)$ subject to the constraint,

$$m_p^2 = \frac{m_0^2(eff) + p_x^2 + p_y^2}{(1-x)} + \frac{m_U^2(eff) + p_x^2 + p_y^2}{x} \qquad (2.3)$$

At this stage, we consider the existence of a set of virtual transitions

$$\left\{ U\uparrow \Leftrightarrow \sum_{q=u,d,s} q\downarrow (\bar{q}\downarrow u\uparrow)_{\pi,K} \right\}_{L_Y=+1} \quad (2.4)$$

that lead to a finite probability of resolving the "massive: constituent quark, $U\uparrow$, into an ensemble of massless quarks and antiquarks. In particular, we focus attention on transitions where the $q\bar{q}$ pair is produced in a $^3P_0$ state with the $\bar{q}\downarrow$ combining with the $u\uparrow$ to form a virtual pion or kaon. As indicated in (2.4), these particular transitions are necessarily $L_y=1$ and describe $A_r$-odd spin-orbit dynamics that contribute to $|M^-|^2$ in (1.3). There are certainly other types of virtual processes involved in the internal dynamics of a proton. However, Reference [23] provides solid phenomenological support for the hypothesis that the virtual processes (2.4) included in the Georgi-Glashow model engender a "pion tornado" that explains crucial aspects of proton structure. To describe the impact of these dynamics on a hard-scattering process, we, for simplicity, set $p_z = p_x = p_y = 0$ so that the constituent U-quark is at rest in the proton CM system, so that

$$P_U^v = (x_0 P^+, \frac{m_U^2(eff)}{x_0 P^+}, 0, 0), \quad (2.5)$$

with $x_0 \cong 1/3$. Take the $q=d$ term in (2.4) and consider the transition $U \Rightarrow d\bar{d}u$ with

$$P_U^v = k_u^v + k_d^v + k_{\bar{d}}^v \quad (2.6)$$

The sketch in Fig. 1 indicates how a positive nonzero expectation value of the operator describing the orbital angular momentum of the u-quark from this configurations

$$L_y^u = (b_z k_x - b_x k_z)^u \quad (2.7)$$

might distribute the expectation value of momentum in the $(b_x, b_z)$ with $b_y \cong 0$ when the massless u-quark is subject to confining boundary conditions.

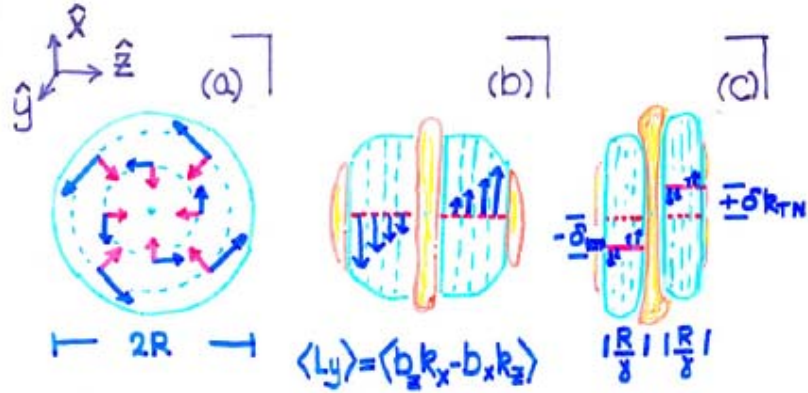

Fig. 1 The sketch in panel (a) indicates orbits in the $(b_x, b_z)$ plane with tangential vectors

showing momentum and inward radial vectors showing the confining forces. The sketch in panel (b) averages over $b_y$ and $b_x$ to show expectation values $\langle k_x \rangle$ in different bins of $b_z$. Panel (c) shows a longitudinal boost with $E = \gamma m_p$ and a separation into two ensembles. The left ensemble with $\langle k_x \rangle = -\delta k_{TN}$ and the right ensemble with $\langle k_x \rangle = +\delta k_{TN}$

It is important to keep in mind that the virtual transitions (2.4) produce expectation values $\langle L_y^u \rangle$ that represent only fractional values of $\hbar$. The orientations of both the momenta and the accelerations $\vec{\alpha} = \partial \vec{k} / \partial t = \partial \vec{k} / \partial \ell$ (where $\ell$ is the path length of the orbit) are important in understanding the kinematics of a scattering event. The projection of the $A_\tau$-odd dynamics into the $(b_x, b_z)$ plane tells only part of the story. Further information is available by restoring the internal dynamics of the constituent quark-diquark system and looking at a projection on the $(b_x, b_y)$ plane as shown in Fig. 2

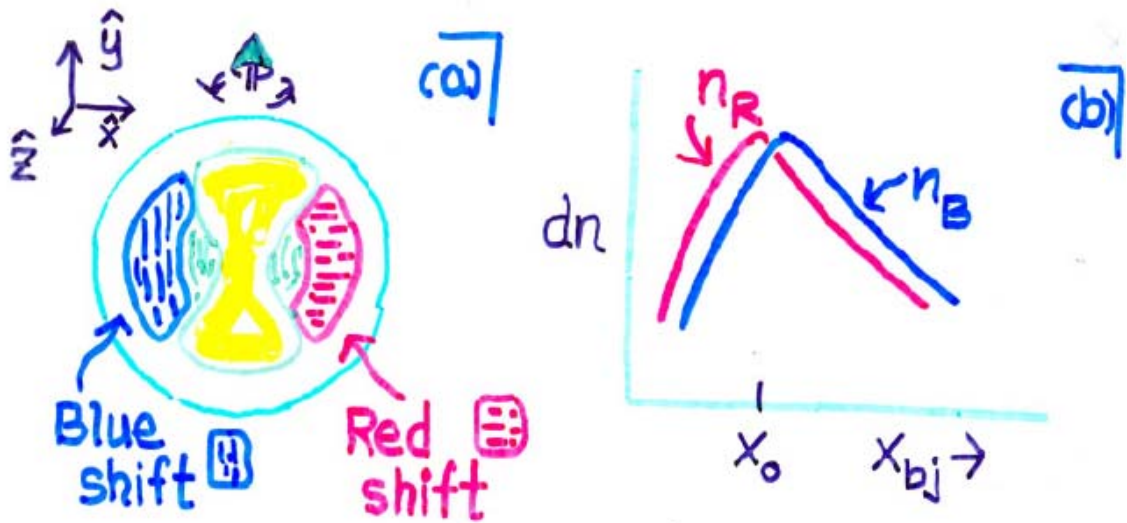

Fig. 2 Projection onto $(b_x, b_y)$ for a system with $\langle L_y^u \rangle \geq 0$ is shown in panel (a). The quark momentum distribution is "blue-shifted" for $b_x \leq 0$ and "red-shifted" for $b_x \geq 0$. Panel (b) shows the distributions: $dn_R / dx$ and $dn_B / dx$ in Bjorken-x for a valence-type quark density distribution where the spin-averaged distribution is peaked at $x = x_0$.

Combining the information contained in these figures, we can see that any u-quark density distribution that represents the difference between the proton spin (and, hence, the constituent U-quark spin) being oriented in the $+\hat{y}$ direction compared to being oriented in the $-\hat{y}$ direction can be defined in terms of 4 distinct ensemble averages. The different ensembles are indicated in the drawing of Fig. 3.

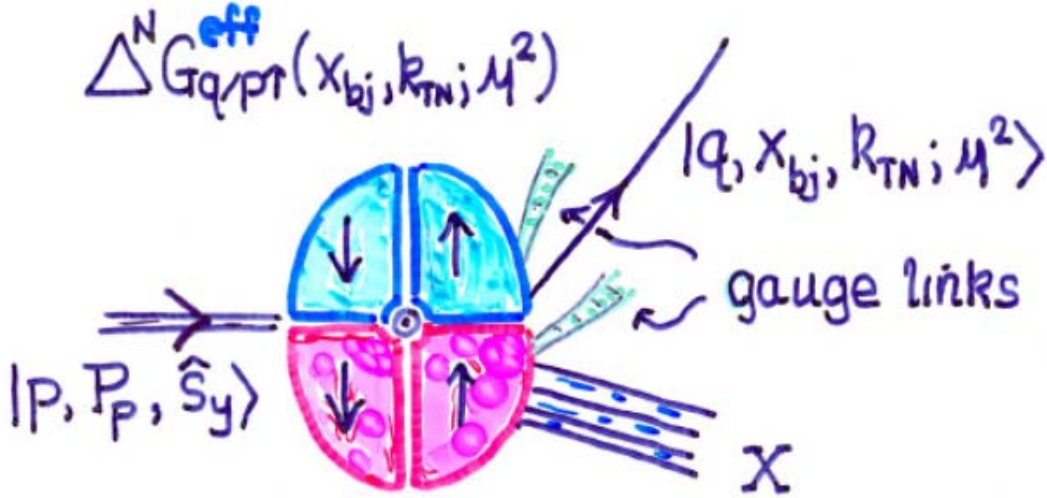

Fig. 3 The "effective" distribution $\Delta^N G^{eff}_{q/p\uparrow}(x, k_{TN}; \mu^2)$ is represented schematically in terms of four distinct ensemble averages describing the correlation in impact parameter space between $\langle L^q_y \rangle$ and the momentum components of an orbiting quark as found in (2.7). The virtual gauge links provide a reminder that, in the absence of nontrivial initial-state or final-stage interactions, the net contribution to a hard-scattering event from these four ensembles will vanish.

The construction of the $A_\tau$-odd effective distribution from the spin-orbit dynamics inherent in the pion tornado of the Georgi-Manohar chiral quark model demonstrates the usefulness of KPR factorization. The further analysis of the virtual transitions in this model presented in Ref. [23] includes phenomenological parameterizations of all the orbital distributions for quarks and antiquarks and of the Boer-Mulders [24] functions for quarks. At the level of virtual corrections defined by (2.4) the Boer-Mulders functions for antiquarks vanish. However, this is a specific feature of the Georgi-Manohar model that may not necessarily be realized in the experimentally-measured distributions.

The isolation of nonperturbative dynamics provided by KPR factorization separates the intrinisic component of orbital distributions representing nuclear structure from the process-dependent dynamical component of orbital distributions. In the parton-model limit rotational invariance requires the contributions from the four sectors to be equal. The kinematically-skewed ensembles constructed above require initial-state or final-state interactions in order to produce non-zero single-spin asymmetries. The process-dependence of orbital distribution is thoroughly discussed in refs. [17,23] within the context of non-perturbative spin-orbit correlations. Depending on color flow, ISI and

FSI involve a combination of shielding effects and deflections involving confining forces. To understand these properties in terms of gauge-invariant local operators requires the gauge link formalism as defined in [4,25]. In this formalism the process dependence is determined by the gauge links that account for Wilson lines in scattering processes involving nonAbelian charges. The sketch of the ensembles representing the $A_\tau$-odd orbital distribution $\Delta^N G^{eff}_{q/p\uparrow}(x, k_{TN}; \mu^2)$ in Fig. 3 includes virtual gauge links to account for this incipient process dependence. In order to explore the connection between initial-state and final-state interactions with non-zero asymmetries we consider defining the asymmetries in terms of spin-directed momentum transfers. The reconciliation between the non-locality involved in the isolation of $A_\tau$-odd dynamics and the gauge-link formalism requires an understanding of the spin-directed momentum transfers involved. Therefore, before turning to a discussion of Collins conjugation, we want to elaborate on some features of single-spin asymmetries that can clarify this concept.

## 3. QCD Evolution and Spin-Directed Momentum Transfers

The most convenient way to parameterize the $A_\tau$-odd dynamics that is isolated by KPR factorization introduces the concept of a spin-directed momentum transfer. For an $A_\tau$-odd distribution function (an orbital distribution or a Boer-Mulders function) this would be a process-dependent transverse momentum of the form $\langle \delta k_{TN}(x, \mu^2) \rangle$ that depends both on the spin-orbit correlations and on the initial-state or final-state interactions that probe the virtual spin-orbit correlations inside the nucleon. For an $A_\tau$-odd fragmentation functions (a Collins function [25] or polarizing fragmentation function [13]) this would be a rank-dependent transverse momentum of the form $\langle \delta k_{TN}(z, \mu^2) \rangle$ that is generated by spin-orbit correlations in the color rearrangement of the fragmentation process. At this point, it is appropriate to concentrate on orbital distributions. The value of the spin-directed momentum transfer approach is that it directs attention to the kinematic variables involved in the underlying spin-orbit dynamics. To see this we can apply the projection operators $\Pi_A^\pm$ to the two independent spin-oriented quark distributions in a proton,

$$\int dk_{TS} G^{eff}_{q/p\uparrow}(x, k_{TN}, k_{TS}; \mu^2) = G^+_{q/p}(x, k_{TN}) + G^-_{q/p}(x, k_{TN})$$
$$\int dk_{TS} G^{eff}_{q/p\downarrow}(x, k_{TN}, k_{TS}; \mu^2) = G^+_{q/p}(x, k_{TN}) - G^-_{q/p}(x, k_{TN}),$$
(3.1)

where we have integrated over $k_{TS}$ to concentrate on the $A_\tau$-odd transverse variable $k_{TN}$ and suppressed the dependence on the factorization scale $\mu^2$. To further simplify the notation we will also suppress the dependence on Bjorken x and write,

$$G^+_{q/p}(k_{TN}) + G^-_{q/p}(k_{TN}) = G^+_{q/p}(k_{TN})(1 + A_N(k_{TN}))$$
$$G^+_{q/p}(k_{TN}) - G^-_{q/p}(k_{TN}) = G^+_{q/p}(k_{TN})(1 - A_N(k_{TN})).$$
(3.2)

At this point, we now define the quantity $\langle \delta k_{TN} \rangle$ by

$$\frac{1}{2}\langle \delta k_{TN} \rangle = \frac{-A_N(k_{TN})G^+_{q/p}(k_{TN})}{\partial G^+_{q/p}(k_{TN})/\partial k_{TN}}$$
(3.3)

and re-express (3.2) in the form:

$$G^{eff}_{q/p\uparrow}(k_{TN}) = G^+_{q/p}(k_{TN} - \frac{1}{2}\langle \delta k_{TN} \rangle)(1 + O\left(\frac{\langle \delta k_{TN} \rangle^2}{m_p^2}\right))$$

$$G^{eff}_{q/p\downarrow}(k_{TN}) = G^+_{q/p}(k_{TN} + \frac{1}{2}\langle \delta k_{TN} \rangle)(1 + O\left(\frac{\langle \delta k_{TN} \rangle^2}{m_p^2}\right)).$$
(3.4)

In these expressions $G^+_{q/p}(k)$ is an even function in k that is sharply peaked around its maximum value of $G^+_{q/p}(0)$. Since the average orbital angular momentum in these transitions is a small fraction of $\hbar$, the ratio $\langle \delta k_{TN} \rangle / m_p$ is necessarily small. Neglecting the terms of $O\left(\frac{\langle \delta k_{TN} \rangle^2}{m_p^2}\right)$ we can re-insert the dependence on x and $\mu^2$ and write,

$$G^{eff}_{q/p\uparrow}(x, k_{TN}; \mu^2) = G^+_{q/p}(x, k_{TN} - \frac{1}{2}\langle \delta k_{TN}(x, \mu^2) \rangle; \mu^2)$$
(3.5)

$$G^{eff}_{q/p\downarrow}(x, k_{TN}; \mu^2) = G^+_{q/p}(x, k_{TN} + \frac{1}{2}\langle \delta k_{TN}(x, \mu^2) \rangle; \mu^2).$$
(3.6)

The shape of the spin-averaged distribution

$$G_{q/p}(x, k_{TN}; \mu^2) = \frac{1}{2}[G^+_{q/p}(x, k_{TN} - \frac{1}{2}\langle \delta k_{TN}(x, \mu^2) \rangle) + G^+_{q/p}(x, k_{TN} + \frac{1}{2}\langle \delta k_{TN}(x, \mu^2) \rangle)]$$ (3.7)

therefore plays a significant role in understanding the behavior of $A_N(x, k_{TN}; \mu^2)$. This is indicated in the drawings of Fig. 4 shown below. It is obvious from the structure of these equations that a small shift on a steeply-falling curve in $k_{TN}$ can produce a large value of the asymmetry $A_N(x, k_{TN}; \mu^2)$ and this connection between large asymmetries and strong dependence on transverse-momentum is verified experimentally. This formulation of single-spin asymmetries serves to demonstrate that it is counterproductive to define $A_N(x, k_{TN}; \mu^2)$ independently from $G^+_{q/p}(x, k_{TN}; \mu^2)$.

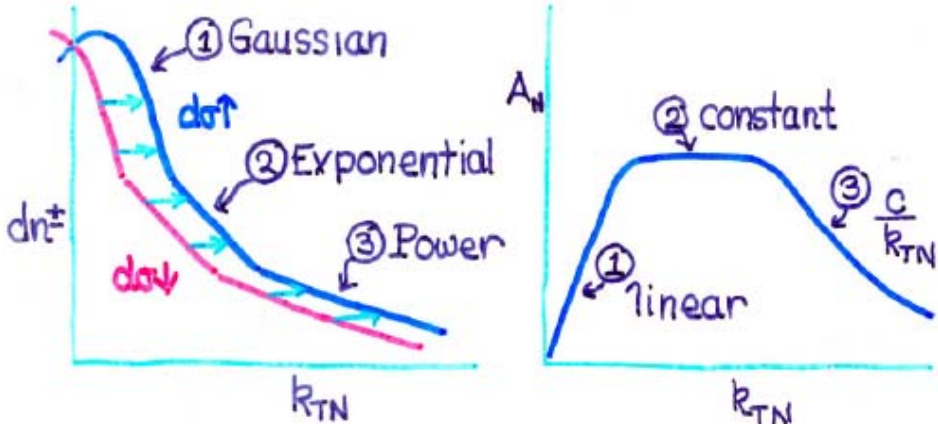

Fig. 4. The sketch in panel (a) shows the functions $G_{q/p\uparrow}(k_{TN})$ and $G_{q/p\downarrow}(k_{TN})$ for different intervals of $k_{TN}$ where the intervals display distinct types of functional dependence. The two curves are shown with a fixed $\langle \delta k_{TN} \rangle$. The sketch in panel (b) shows the corresponding behavior of $A_N(k_{TN})$ in these regions.

This figure illustrates the close connection between the functional form of $G^+_{q/p}(x, k_{TN})$ at fixed value of Bjorken-x and the corresponding value of $A_N(x, k_{TN})$ based on a constant value of $\langle \delta k_{TN}(x, \mu^2) \rangle$. For a gaussian behavior at small $k_{TN}$,

$$G^+_{q/p}(k_{TN})_{k_{TN} \leq m_p} = a_g \exp[-b_1^2 k_{TN}^2], \tag{3.8}$$

we get the asymmetry

$$A_N(k_{TN})_{k_{TN} \leq m_p} = \tanh[b_1^2 \langle \delta k_{TN} \rangle k_{TN}]. \tag{3.9}$$

Then, for intermediate values of $k_{TN}$ where $G^+_{q/p}(x, k_{TN})$ displays an exponential behavior in $k_{TN}$,

$$G^+_{q/p}(k_{TN})_{k_{TN} \approx m_p} = a_m \exp[-b_2 k_{TN}], \tag{3.10}$$

we get a constant behavior in $k_{TN}$ with a value

$$A_N(k_{TN})_{k_{TN} \approx m_p} = \tanh[b_2 \langle \delta k_{TN} \rangle]. \tag{3.11}$$

In each of these regions, we can determine the value of $\langle \delta k_{TN}(x, \mu^2) \rangle$ by measuring the slope of $G^+_{q/p}(k_{TN})$. Finally, at large values of $k_{TN}$ where KPR factorization ensures that the formalism agrees with the prediction of the appropriate higher-twist operators,

$$G^+_{q/p}(k_{TN})_{k_{TN} \gg m_p} = a_3 \left(\frac{k_{TN}}{\mu}\right)^{-N_{eff}}, \tag{3.12}$$

we get a fall-off of the asymmetry with the form,

$$A_N(k_{TN})_{k_{TN} \gg m_p} = \frac{1}{2} N_{eff} \frac{\langle \delta k_{TN} \rangle}{k_{TN}} \tag{3.13}$$

An important feature of the spin-directed momentum transfer is that, for distributions obeying TMD factorization, we can show that $\langle \delta k_{TN}(x,\mu^2) \rangle$ does not evolve under TMD evolution. This is indicated in the drawings shown in Fig. 5 and follows from the isolation of $A_\tau$-odd dynamics implied by KPR factorization.

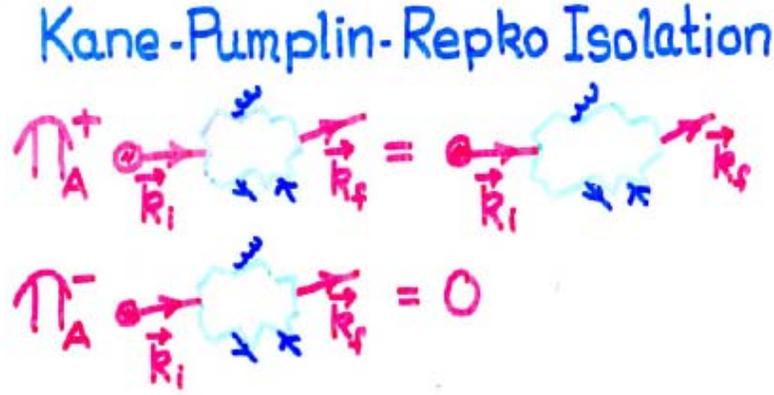

Fig. 5 The operators $\Pi_A^\pm$ acting on diagrams in perturbative QCD.

Since there is no preferred spin orientation in QCD perturbation theory, if we evolve the spin-dependent momentum transfer from a typical hadronic scale, $\mu^2$, with $\langle \delta k_{TN}(x,\mu^2) \rangle = \delta$, such as

$$\vec{k}_i(\mu^2) \uparrow = (xP, +\delta/2, 0)$$
$$\vec{k}_i(\mu^2) \downarrow = (xP, -\delta/2, 0), \tag{3.14}$$

the perturbative mechanisms describing the resolution of the emissions shown in Fig. 5 at a hard-scattering scale $Q^2 \gg \mu^2$ would lead to,

$$\vec{k}_f(Q^2) \uparrow = (xP\cos\theta(Q^2), xP\sin\theta(Q^2)\sin\phi(Q^2) + \delta/2, xP\sin\theta(Q^2)\cos\phi(Q^2))$$
$$\vec{k}_f(Q^2) \downarrow = (xP\cos\theta(Q^2), xP\sin\theta(Q^2)\sin\phi(Q^2) - \delta/2, xP\sin\theta(Q^2)\cos\phi(Q^2)). \tag{3.15}$$

Because $\langle \sin\phi(Q^2) \rangle = \langle \cos\phi(Q^2) \rangle = 0$, there would be no change in the mean value $\langle \delta k_{TN}(x,\mu^2) \rangle = \langle \delta k_{TN}(x,Q^2) \rangle = \delta$ although both Bjorken x and the mean-squared values $\langle k_{TN}^2 + k_{TN}^2 \rangle = \langle k_T^2 \rangle = x^2 P^2 \sin^2\theta(Q^2)$ would evolve with $Q^2$ leading to the improved resolution at higher $Q^2$. The spin-directed momentum shift $\langle \delta k_{TN}(x,\mu^2) \rangle$ provides a rigorous definition of a spin asymmetry within an orbital distribution resulting from the sampling process that exposes the non-local spin-orbit dynamics. It necessarily depends upon Bjorken x and on a nonperturbative scale, $\mu^2$, with magnitude of order of a constituent quark mass squared. The invariance of $\langle \delta k_{TN}(x,\mu^2) \rangle$ under TMD evolution implies that the evolution of $A_N(x, k_{TN}; Q^2)$ is closely constrained by the TMD evolution of $G^+(x, k_{TN}; Q^2)$ calibrated by (3.5) and (3.6). An illustration of this constraint is shown in Fig. 6. The phenomenological studies of TMD evolution for $\Delta^N G(x, k_{TN}; Q^2)$ by Aybat, Rogers and collaboraters [26], by Anselmino et al. [27] and by Boer [28] help quantify the behavior indicated in these sketches.

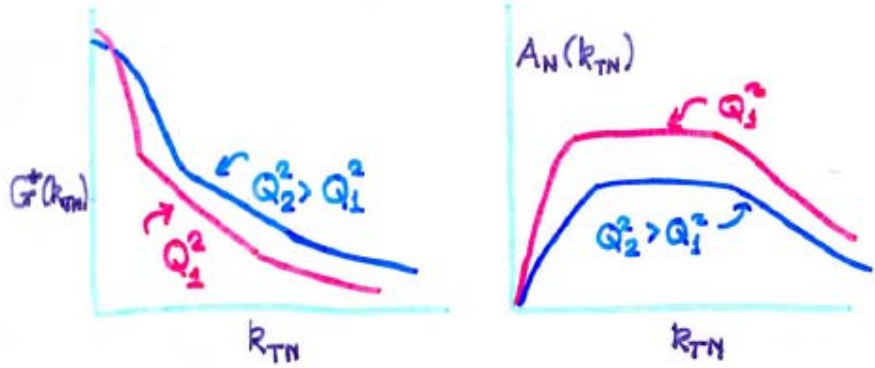

Fig. 6 The TMD evolution of $A_N(x, k_{TN}; Q^2)$ resulting from a fixed $\langle \delta k_{TN}(x, \mu^2) \rangle$ combined with the evolution of $G^+_{q/p}(x, k_{TN}; Q^2)$.

We have seen that the dynamical mechanisms leading to single-spin asymmetries can be parameterized by the spin-directed momentum transfer $\langle \delta k_{TN}(x, \mu^2) \rangle$. The fact that the momentum shift does not change under TMD evolution provides an important calibration by relating the change in slope of the spin-averaged cross section $G^+_{q/p}(x, k_{TN}; Q^2)$ to the change in magnitude of $A_N(x, k_{TN}; Q^2)$. In addition, the quantity $\langle \delta k_{TN}(x, \mu^2) \rangle$ provides a well-defined measure of the process dependence that can be reliably calculated with methods of quantum mechanics. This formalism, thus, can be combined with KPR factorization both to aid in the understanding of TMD evolution for $A_\tau$-odd distribution functions and to fully understand the underlying mechanisms leading to transverse single-spin asymmetries involving these functions. The relationship between the spin-directed momentum transfer and the initial-state and final-state interactions that expose the spin-orbit correlations found above can best described in terms of a simple comparison between DY and SIDIS. Consideration of the dynamical processes involved in this comparison leads to the topic of Collins conjugation.

## 4. Collins conjugation: the comparison of transverse single-spin asymmetries in semi-inclusive deep inelastic scattering and the Drell-Yan process.

The study of QCD has benefited significantly from detailed phenomenological comparisons involving the processes: $e^+e^- \Rightarrow hadrons$, semi-inclusive deep-inelastic scattering (SIDIS) $e^-p \Rightarrow e^-hX$, and the Drell-Yan (DY) process $ph \Rightarrow l^+l^-X$. [29] The current investigation of TMD's continues this saga by introduction the transverse momentum dependence of hadronic distribution and fragmentation functions. The properties of TMD factorization and TMD evolution play an important role in the experiments being planned [2,3] to make precision tests of Collins conjugation as described by Eq. (1.1). The validity of TMD factorization has been demonstrated [6,7] for the transverse-momentum dependent distribution functions and fragmentation functions for the three processes identified above. The strikingly nonintuitive prediction, (1.1), found by John Collins [4] relating to orbital distribution that leads to a single-spin asymmetry in the Drell-Yan process with that observed in SIDIS [30,31] not only provides a stringent test of the underlying foundations of TMD factorization but also can be seen to be a fundamental confirmation of the gauge principle in QCD. To see this, we can call on the results above to rewrite (1.1) in the framework of a spin-directed momentum,

$$\left\langle \delta k_{TN}(x,\mu^2) \right\rangle_{DY} = -\left\langle \delta k_{TN}(x,\mu^2) \right\rangle_{SIDIS}. \qquad (4.1)$$

This spin-directed momentum formulation allows the comparison of observables at different values of $Q^2$ and removes uncertainties involved with TMD evolution. Precision measurements are required to confirm both the magnitude and the sign in the comparison. The original derivation of Collins conjugation [4] requires the application of time-reversal or charge conjugation to relate the gauge-link formulation of the orbital distribution describing the two processes. This derivation has been confirmed in specific model calculations [21,32] based on spectator models and Feynman diagrams. In addition, the observed sign for the SIDIS asymmetry and the change in sign for DY is supported by semi-classical arguments [33] based on confining forces that model the initial-state (DY) interactions and the final-state (SIDIS) interactions involved in the two processes.

The spin-directed momentum approach combined with KPR factorization allows for a more direct understanding of Collins conjugation in terms of Wilson loops [34,17] giving the space-time description of the color flow associated with the two hard processes. We restrict attention to regions of Bjorken x where the density distributions fall with increasing x so that the "blue-shifted" momentum ensemble of Fig. 2 contributes preferentially to each of the cross sections. This requires that the expectation value for $\left\langle \delta k_{TN}(x,\mu^2) \right\rangle$ be dominated by contributions from spatial regions with $\left\langle b_x \right\rangle \geq 0$. As indicated in Fig. 7, we can consider a projection in the $(b_y, b_z)$ plane and draw Wilson loops for each of the processes.

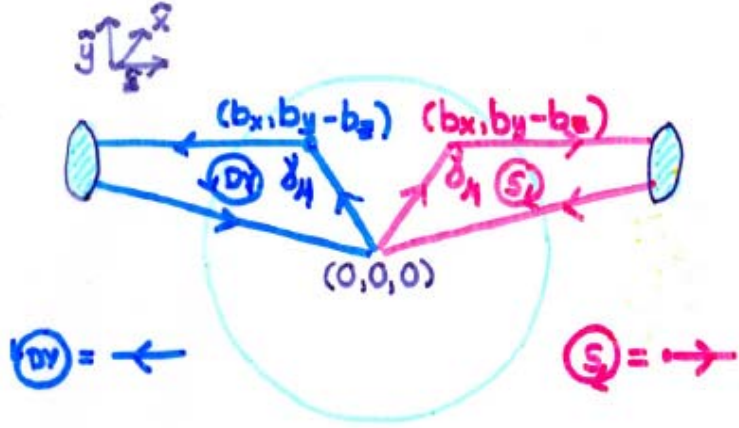

Fig. 7 Wilson loops for DY and SIDIS asymmetries.

Rotational invariance around the y-axis in this figure suggests we pair vertex points $(b_x, b_y, b_z)_{SIDIS}$ with $(b_x, b_y, -b_z)_{DY}$ for the hard scattering location in the two processes. The causal path for each of the path integrations shown in these loops can be interpreted as starting at the origin at time $t = T_o$ and ending at the origin at time $t = T_o + \Delta t$ along with an additional link in the timelike direction at the origin to form a closed loop in Minkowski space. The non-Abelian path integrations are conveniently evaluated in the radial coordinate gauge $\vec{A}_a(\vec{r}, t) \cdot \hat{r} = 0,$ where the confining force in the rest frame of the rotating quark involves only and electrical component. In this gauge, only the "horizontal" segments contribute to the path-ordered integrations. For each location of the two vertices, we can apply the nonAbelian Stokes theorem yielding the deflection,

$$\delta k_{TN}(b_x, b_y, b_z)_{SIDIS} = -\delta k_{TN}(b_x, b_y, -b_z)_{DY} \qquad (4.2)$$

and, averaging over all possible locations for the vertices we recover Eq. (4.1) for the spin-directed momentum transfer of the two related processes. The geometrical symmetry between the two Wilson loops allows the result to be verified in many gauges. The Wilson loops shown in Fig. 7 have also been estimated numerically. This was done in an important lattice Monte Carlo simulation from Musch, Hagler, Engelhart, Nagle and Shafer [35] that also independently confirms the validity of Collins conjugation. The sketch in Fig. 8 shows the $\delta k_{TN}$ "shift" for the quantum numbers representing "u-d" quarks based on a figure from this paper.

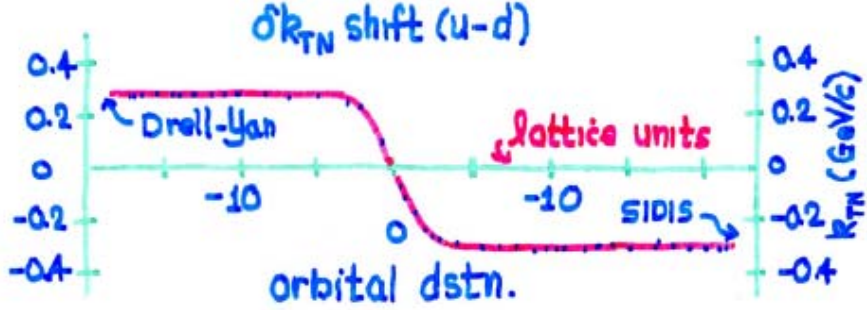

Fig. 8 The $\delta k_{TN}$ shift as a function of lattice spacing number for a "u-d" quark orbital distribution from the calculations of Ref. [35].

The calculations in this paper are done in Euclidean space-time with gauge configurations far from the chiral limit. The connection to the experimental observables is explained more fully in the paper but the lattice calculation supports the observation first expressed by Piljman [34] that Collins conjugation can be placed in formal analogy to the Aharanov Bohm asymmetry [36] in QCD. However, where the Aharanov Bohm asymmetry tests the gauge formulation of QED in regions where the electromagnetic field-strength tensor vanished, Collins conjugation tests the gauge formulation of QCD at scales where quarks and gluons may not fully describe the dynamical degrees of freedom of hadronic physics.

If we return to the Georgi-Manohar model of Sec. 2 and assume that <u>not all</u> of the forces associated with the pion tornado generated in this model are adequately described by the Wilson loops of Figs. 7-8, we conclude that the strict equality of (4.2) can be modified by model-dependent corrections,

$$\langle \delta k_{TN}(x,\mu^2) \rangle_{DY} = -\langle \delta k_{TN}(x,\mu^2) \rangle_{SIDIS} [1+\varepsilon(x,\mu^2)] \qquad (4.3)$$

with the correction term,

$$\varepsilon(x,\mu^2) = \sum_n b_n(x,\mu^2)\left(\frac{\delta k_{TN}}{m_\pi}\right)^n + \sum_n c_n(x,\mu^2)\left(\frac{m_\pi}{m_Q}\right)^2 + \sum_n d_n(x,\mu^2)\left(\frac{m_Q}{m_p}\right)^2 \qquad (4.4)$$

Within the framework of the Georgi-Manohar model, the individual corrections in these sums may each be small for many reasons in an arbitrary "effective field theory" estimate of the virtual fluctuations. However, they must all vanish identically under the hypothesis that QCD can be formulated as a gauge theory so that the all spin-oriented momentum shifts are generated by the Wilson loops shown in Fig. 7. The range of scales associated with these issues are indicated in Fig. 9

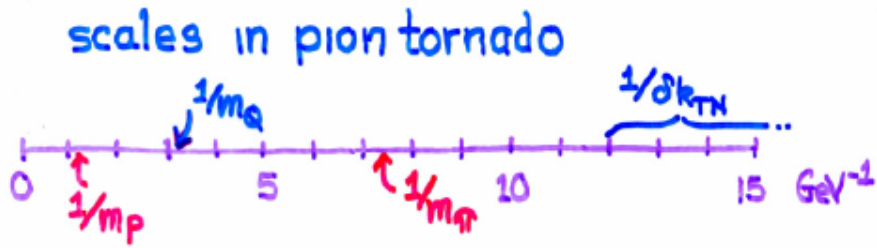

Fig. 9 Using the framework of the Georgi-Manohar chiral quark model as a basis for an effective field theory, ratios of the relevant mass scales $m_p \geq m_Q \geq m_\pi \approx \delta k_{TN}$ can lead to corrections in the Collins conjugation relation as indicated in (4.4). These scales are indicated as distances in impact parameter space.

     This example serves to quantify the requirement for precision tests of Collins conjugation. The isolation of nonperturbative dynamics implied by KPR factorization has proved to be an important phenomenological tool. The lattice simulations of Musch et al. [34] strongly support the theoretical basis for this isolation. The theoretical basis for Collins conjugation is both consistent and compelling. However, it cannot be considered more than an interesting prediction unless it is verified experimentally. Therefore, stringent experimental verification of (4.1) or (4.4) is required to validate the idea that the underlying dynamical mechanisms leading to single-spin asymmetries can be calculated with a gauge field theory. The refutation of Collins conjugation would indicate that there are important aspects of hadronic physics that cannot be quantitatively described by QCD. Thus, the E-1027 experiment [2] has an opportunity to do something very important that can also lead to further discoveries.

The author is grateful to Andreas Metz for suggesting the topic of these presentations and to John Collins, Gary Goldstein, Simonetta Liutti, and Daniel Boer for invaluable comments and valuable guidance.

# REFERENCES:

1. D. Sivers, Nuovo Cim. **35C N.2**; 171 (20120
2. L.D. Eisenhower et al., (SeaQuest collaboration) "Polarized Drell-Yan Measurements with the Fermilab Main Injector" E-1027 proposal
3. J. Dudek et al. "Physics Opportunities with the 12 GeV Upgrade at Jefferson Lab" arXiv:1208.1244 [hep-ex]
4. J. Collins, Phys. Lett. **B536**, 43 (2002)
5. NSAC Long Range Plan (2007), http://science,energy,gov/np/nsac
6. J.C. Collins, D.E. Soper, G. Sterman, in *Perturbative Quantum Chromodynamics* (A.H. Mueller, Editor, World Scientific, Singapore, 1989) ; J. C. Collins , *Foundations of Perturbative QCD* (Cambridge Univ. Press, UK, 2011).
7. M. Echevarria, A. Idilbi, I. Scememi, JHEP07 (2012), 002; arXiv:1205.3892;arXiv:1211.1947 (2012)
8. G. Kane, J. Pumplin and W. Repko, Phys. Rev. Lett. **41**, 1689 (1978).
9. A. Bacchetta, U.D'Alesio, M. Diehl and C.A. Miller, Phys. Rev. **D70**, 117504 (2004).
10. See, for example, A. Krisch, Sci. Am. **257**, 42 (1987).
11. A.V. Efremov and O.V. Teryaev, Sov. J. Nucl. Phys. **36**, 140 (1982); Phys. Lett. **B150**, 383 (1985).
12. D. Sivers, Phys. Rev. **D41**, 82 (1990).
13. D. Sivers, Phys. Rev. **D43**, 261 (1991).
14. J. Qiu and G. Sterman, Phys. Rev. Lett. **67**, 2264 (1991); Nucl. Phys. **B378**, 52 (1992).
15. A. Metz and D.Pitonyak, arXiv:1212.5037
16. W.G. Dharmaratna and G. Goldstein, Phys. Rev. **D41**, 1731 (1990); **D53**, 1073 (1996).
17. D. Sivers, Phys. Rev. **D74**, 094008 (2006)
18. Y.S. Derbenev and A. Kondratenko, Part. Accel. **8**, 115, (1978).
19. R. Jaffe, in Proceedings of the Workshop on Deep Inelastic Scattering off Polarized Targets, DESY-Zeuthen, Germany, 1997, edited by J. Blumlein et al.,p. 167.
20. D. Sivers, in Proceedings of the Workshop on Deep Inelastic Scattering off Polarized Targets, DESY-Zeuthen, Germany, 1997, edited by J. Blumlein et al., p 383.
21. S. Brodsky, D. Hwang, Y. Kovchegov, I Schmidt and M. Sievert, arXiv:1304.5237 (2013).
22. A. Manohar and H. Georgi, Nucl Phys. **B234**, 189 (1984).
23. D. Sivers, "Chiral Mechanisms Leading to Orbital Structures in the Nucleon" arXiv:0704.1791 (unpublished)
24. D. Boer and P. J. Mulders, Phys. Rev. **D57**, 5780 (1998).

25. J. Collins, Nucl. Phys. **B396**, 161 (1993).
26. S. Aybat and T. Rogers, Phys. Rev. **D83**,114042 (2011); S. Aybat, J. Collins, J.-W. Qiu and T. Rogers, Phys Rev. **D85**, 034043 (2012).
27. Anselmino, M. Boglione, U. D'Alesio, S. Mulis, F. Murgia, A. Prokudin, arXiv:1304.7691 (2013)
28. D. Boer, arXiv:1304.5387
29. A.S. Kronfield and C. Quigg, "Resource Letter: Quantum Chromodynamics" Am. J. Phys. **78**, 1081=1116 (2010).
30. A. Airapetian et al. (Hermes collaboration) Phys Rev. Lett. **103**, 162002 (2009).
31. M. Alekseev et al. (Compass collaboration) Phys. Lett. **B692**, 240-246 (2010).
32. S. Brodsky, D.S. Hwang and I. Schmidt, Phys. Lett. **B530**, 99 (2002); L. Gamberg, G.R. Goldstein and K.A. Orgavesyan, Phys. Rev. **D67**, 071504 (2003)
33. M. Burkardt, Phys. Rev. **D62**, 071503 (2000).
34. F. Pijlman, Few Body Syst. **36**, 209 (2005); Ph.D. thesis arXiv:hep-ph/0604226 (2006).
35. B. Musch, P. Hagler, M. Engelhardt, J. Negele, and A. Schafer, Phys Rev. **D85**, 094510 (2012)
36. Y. Aharanov and D. Bohm, Phys. Rev. **115** , 485 (1959); **123**, 1511 (1961).